# Tokens All the Way Down:

# A Money View of Decentralized Finance

Wenbin Wu[1][1]

[1]*Cambridge Judge Business School, University of Cambridge, UK*

Corresponding author: w.wu@jbs.cam.ac.uk

**Abstract**

In traditional banking, repeated deposit-and-lend cycles let a single dollar of reserves support multiple dollars of claims. Decentralized finance produces an analogous structure with tokens. Constructing a Token Graph of 10,200 tokens across 200 blockchains, this paper maps the resulting hierarchy and shows that, by late 2025, each dollar of base assets supports $4.7 of total claims. An embedded yield correction disentangles two channels that raw data conflates: a *compositional channel*, where lending protocols concentrate in deeper tiers and mechanically raise average yields; and a *liquidity channel*, where each derivation step reduces secondary-market depth and depresses yields in liquidity-sensitive pools. The liquidity channel concentrates in DEX pools and vanishes in lending pools. A yield decomposition shows that the tier gradient operates entirely through fundamental protocol yields, not incentive-token emissions; quantile regressions reveal that the structural associations concentrate in the upper tail of the yield distribution, with near-zero effects at the median. These findings reframe DeFi's "double counting" as a structural risk question and identify liquidity fragmentation as the primary mechanism associated with yield variation across the token hierarchy.

[1]The author thanks Rashad Ahmed, Nicola Gennaioli, Campbell Harvey, Simon Mayer, Dirk Niepelt, and Andreas Park for helpful comments and suggestions, and Patrick Scott and the DefiLlama team for data access. All errors are the author's own.

# 1 Introduction

In traditional banking, a single dollar of reserves supports multiple dollars of deposits through repeated cycles of lending and redeposit, a process known as the money multiplier. Decentralized finance produces an analogous structure with tokens instead of deposits. When a user deposits ETH into a staking protocol and receives stETH, then deposits that stETH into a lending protocol and receives astETH, the result is a second-order derivative whose value depends on every protocol in the chain. As of late 2025, this process extends across thousands of protocols and over 200 blockchains (DefiLlama 2024), locking over $170 billion in total value (EtherScan 2024).

This paper maps the layering. Constructing a Token Graph of approximately 10,200 tokens, the analysis reveals DeFi as a credit hierarchy where each dollar of base assets supports $4.7 of total claims, a "layering multiplier" analogous to the traditional money multiplier. A token's position in this hierarchy predicts its yield, but only after disentangling two opposing channels. Reported yields *rise* with tier depth because lending protocols, which pay higher interest rates, concentrate in deeper tiers: a compositional effect. Yet after correcting for this composition, yields *fall* with each derivation step from base assets: a structural distance discount. This discount vanishes entirely in lending pools and concentrates in DEX pools, identifying liquidity fragmentation as the primary mechanism.

The recursive nesting of claims is not unique to DeFi. The "Money View" framework (Mehrling 2013) describes traditional finance as a hierarchy: central bank reserves at the top, bank deposits below, and successive layers of credit instruments, each a promise to pay in terms of the tier above. Banks profit from the spread between tiers (Mehrling 2011), and Pozsar (2014)'s shadow-banking hierarchy extends the same logic to non-bank intermediaries. In crypto, Gorton and Zhang (2023) draw an analogy with Free Banking Era banknotes, Anadu et al. (2023) document stablecoin runs, and Olk and Miebs (2025) characterize on-chain credit creation as a form of shadow banking.

A growing literature addresses the measurement challenges of derivative layering. Luo et al. (2024) introduce Total Value Redeemable, distinguishing "plain" from "derivative" tokens; Saggese et al. (2025) propose verifiable TVL; Brigida (2025) document double-counting limitations. These approaches treat derivative layering as measurement noise to be filtered out. The present paper takes the opposite view: each additional tier represents one more round of "deposit, receive receipt, redeposit elsewhere," and this layering *is* the credit structure itself, just as the money multiplier in banking describes how credit expands through intermediation chains. Aldasoro et al. (2023) apply the Money View to stablecoins, modeling them as "crypto Eurodollars" whose par settlement depends on liquidity rather than solvency; the logic extends here from stablecoins to the full DeFi ecosystem.

A complementary strand organizes DeFi by protocol function. Schär (2021) and Auer et al. (2024) propose protocol-layer taxonomies; Werner et al. (2022) surveys composability risks; and Castillo León and Lehar (2026) provide a comprehensive empirical review. Crisis studies document the consequences: contagion through the Terra ecosystem (Badev and Watsky 2023, Briola et al. 2023, Liu et al. 2023), deleveraging spirals from collateral liquidations (Perez et al. 2021, Klages-Mundt and Minca 2022), and the cascading failures analyzed by Gudgeon et al. (2020). More recently, Lee (2026) formalize layered DeFi as a "waterfall of externalities" generating systemic illiquidity, and Heimbach and Huang (2024) measure wallet-level leverage at roughly twice the collateral value. Tovanich et al. (2025) apply DebtRank to lending networks, and Zhang et al. (2026) analyze TVL-based fragility propagation. The present paper complements these protocol-level and wallet-level perspectives by organizing *tokens* by their position in the credit chain, a shift in unit of analysis that makes the hierarchical nesting composability produces explicit and quantifiable.

On DeFi yields, Augustin et al. (2022) document "reaching for yield" reminiscent of traditional credit markets, Cornelli et al. (2024) show that DeFi rates co-move with Treasury yields, and Kitzler et al. (2023), Wu et al. (2025), and Shu et al. (2026) construct inter-protocol exposure graphs. No existing study decomposes DeFi yields by hierarchical position; this paper bridges the yield and network literatures by showing that a token's structural distance from base assets robustly predicts pool-level yield, providing a structural interpretation for the "reaching for yield" phenomenon.

Three contributions follow. *Conceptually*, the Money View extends from stablecoins (Aldasoro et al. 2023) to the full DeFi ecosystem. The Token Graph makes the layered credit structure explicit, providing a macro-prudential metric, the layering multiplier, that tracks system-wide leverage in real time. *Methodologically*, a balance-sheet approach discovers token hierarchies automatically from pool-level data, treating every DeFi protocol as a financial intermediary holding assets and issuing liabilities. *Empirically*, the hierarchy is traced through major disruptions, and an embedded yield correction disentangles two channels previously conflated in raw data: a compositional tier premium driven by the changing protocol mix, and a structural distance discount that survives correction, concentrates in liquidity-sensitive pool types, and widens during crises.

Section 2 develops the conceptual framework, Section 3 describes the data and methodology, and Section 4 and Section 5 present results and discussion.

# 2 Conceptual Framework

## 2.1 DeFi Protocols as Financial Intermediaries

DeFi protocols function as financial intermediaries that accept tokens as deposits and issue new tokens as claims against those deposits, mirroring the basic operation of traditional banking. A bank accepts cash and issues a deposit receipt; Aave accepts WETH and issues aWETH. The fundamental operation, accepting one asset and issuing a claim on it, creates the tiered credit relationships that constitute a monetary hierarchy. Formal definitions are provided in Section A.

Each liquidity pool corresponds to one such transformation: a pool record specifies input tokens, output tokens, and total value locked, as formalized in Equation 3. The directed edges in the token graph derive from these pool-level observations. As Fig. 1 illustrates, when a user deposits USDC into Aave, the protocol issues aUSDC as a receipt token, defining a directed edge USDC $\rightarrow$ aUSDC.

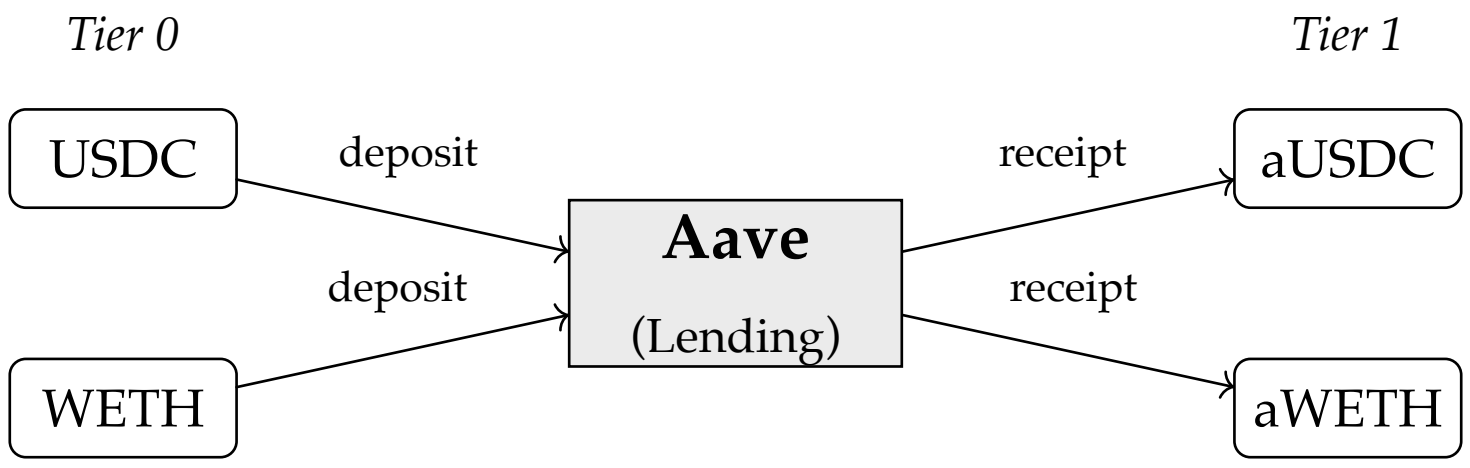

**Fig. 1** | Token transformation in a lending protocol. The protocol accepts Tier 0 tokens as deposits and issues Tier 1 receipt tokens. Each pool represents one such transformation relationship.

## 2.2 The Token Hierarchy

These intermediation relationships give rise to a layered structure. DeFi tokens occupy distinct tiers in a credit hierarchy analogous to the money hierarchy (Mehrling 2013), with each tier defined by its distance from base assets. Tier 0 contains tokens exogenous to DeFi's on-chain ecosystem: native blockchain assets such as ETH, BTC, and SOL, and fiat-backed stablecoins such as USDC and USDT. Stablecoins are claims on off-chain reserves, but within the on-chain topology no protocol creates them from a more primitive token; they function as base money for the dollar-denominated portion of DeFi, much as Eurodollars function as base money in offshore markets despite being claims on onshore USD (Aldasoro et al. 2023).

Tier 1 comprises tokens directly derived from Tier 0: wrapped tokens such as WETH, liquid staking tokens such as stETH, and CDP-minted stablecoins such as DAI. Tier 2 consists of tokens issued by protocols accepting Tier 1 inputs, including lending receipts such as aUSDC and restaking derivatives such as weETH. Higher tiers contain tokens built on Tier 2, including yield aggregator receipts and structured products. Table 1 provides examples at each tier. Formally, a token's tier is the shortest path from any Tier 0 asset, as Equation 5 defines; risk is cumulative, since a Tier 3 token inherits the risks of every protocol in its derivation chain.

| Tier | Token | Description | TVL ($B) |
|---|---|---|---|
| 0 | ETH | Native Ethereum | 40.5 |
| 0 | USDC | USD Coin | 15.9 |
| 0 | USDT | Tether | 6.0 |
| 0 | BTC | Native Bitcoin | 0.6 |
| 0 | SOL | Native Solana | 0.3 |
| 1 | WETH | Wrapped ETH | 44.8 |
| 1 | stETH | Lido Staked ETH | 29.0 |
| 1 | wstETH | Wrapped stETH | 9.2 |
| 1 | WBTC | Wrapped BTC | 7.0 |
| 1 | DAI | MakerDAO Stablecoin | 3.2 |
| 2 | weETH | Wrapped eETH | 15.4 |
| 2 | aWBTC | Aave WBTC | 4.3 |
| 2 | aUSDC | Aave USDC | 3.8 |
| 2 | cWETH | Compound WETH | 1.2 |
| 3 | aweETH | Aave weETH | 6.4 |
| 3 | awstETH | Aave wstETH | 5.3 |
| 3 | spwstETH | Spark wstETH | 2.1 |
| 4 | aEzETH | Aave ezETH | 0.4 |
| 4 | aTETH | Aave tETH | 0.2 |

**Table 1** | Example Tokens by Hierarchy Tier (TVL as of Q4 2025).

## 2.3 Tier Transitions and Credit Creation

The hierarchy classifies three economically distinct tier transitions, each carrying different risk implications. *Format conversion* such as ETH → WETH creates no new purchasing power. *Claim creation via staking* such as ETH → stETH expands the balance sheet; the new claim circulates freely as collateral while the underlying remains locked. *Claim creation via lending* such as WETH → aWETH takes expansion further: the depositor holds the receipt while a borrower obtains the input to deploy elsewhere, supporting two simultaneous circulating claims. The 2024–2025 landscape has introduced additional mechanisms, including restaking, yield tokenization, and delta-neutral synthesis, that the graph-theoretic tier assignment handles correctly, though tokens at the same tier may carry qualitatively different risk profiles.

The traditional money multiplier analogy applies most directly to lending-driven transitions. The layering multiplier, defined as total mapped TVL divided by Tier 0 TVL as formalized in Equation 6, measures how many dollars of claims exist per dollar of base assets. This metric is broader than the banking money multiplier: it captures all forms of derivative layering, and its decomposition in Section 4 distinguishes contributions of wrapping, staking, and lending.

## 2.4 Hypothesis Development

If DeFi is organized as a credit hierarchy, the Money View generates testable predictions about yields. Each additional tier introduces incremental smart contract risk, liquidity risk, and counterparty risk, so theory predicts that yields compensate for accumulated exposures. Two channels could link hierarchical position to yields, and they need not operate in the same direction. First, a *compositional gradient*: the Money View predicts that staking dominates lower tiers while lending dominates upper tiers, so yields may vary with the changing protocol mix at each depth. Second, *structural distance* from base assets may independently reduce yields, since more deeply nested tokens attract fewer borrowers, lowering utilization rates and consequently algorithmically determined lending rates. These channels can oppose each other: the compositional gradient predicts higher yields at deeper tiers, while structural distance may predict lower yields.

**Hypothesis 1 (Yield–Hierarchy Association):** *Yields vary systematically with hierarchical position, after controlling for pool characteristics and market-wide yield fluctuations.* H1a: the tier-yield gradient operates through the changing protocol mix across hierarchy depths. H1b: conditional on protocol type, derivation hops from base assets independently predict yields.

**Hypothesis 2 (Countercyclical Premium):** *The tier yield gradient increases during market stress, consistent with repricing of upstream dependency risk.* During crises, participants demand higher compensation for holding claims that depend on multiple upstream protocols, or sell those claims to move toward base money (Mehrling 2011).

# 3 Data and Methodology

## 3.1 Data Sources

The analysis draws on DeFi data from DefiLlama (DefiLlama 2024), covering over 7,000 protocols across more than 200 blockchains. The dataset includes approximately 18,000 liquidity pools with input and output tokens, TVL in USD, annualized percentage yield, and blockchain and protocol identifiers. The complete tier assignment algorithm is provided in Section A.6.

DefiLlama aggregates TVL through protocol-specific "adapters" contributed by protocol teams, introducing reporting heterogeneity. Saggese et al. (2025) estimate that under half of aggregate DeFi TVL can be independently verified on-chain. The tier assignment depends primarily on the *topology* of token relationships rather than TVL magnitudes, providing partial insulation from reporting errors.

For lending protocols such as Aave, Compound, and Morpho, receipt tokens are not always explicitly recorded. Receipt token symbols are inferred using protocol-specific naming conventions, an essential step for mapping token transformation relationships in lending markets. Cross-chain token identity is resolved by matching on symbol and canonical contract address where available; tokens that share a symbol across chains but represent economically distinct assets, such as governance tokens with chain-specific deployments, are treated as separate nodes.

## 3.2 Token Graph Construction

A directed graph $G_{\text{full}} = (\mathcal{X}, \mathcal{E}_{\text{full}})$, formalized in Equation 4, represents all token transformations: vertices are tokens and edges are protocol-mediated transformations carrying metadata on protocol, blockchain, and TVL. The resulting graph contains approximately 10,200 nodes and 12,100 edges. The graph is predominantly acyclic, with 79 cycles involving negligible TVL after excluding the trivial ETH $\leftrightarrow$ WETH wrapper; cycles are resolved as described in Section A.7.

## 3.3 Tier Assignment

Rather than manually specifying base tokens, the hierarchy is discovered from graph topology using a balance-sheet approach grounded in the Money View: each non-trading DeFi protocol holds assets and issues liabilities, defining a derivation relationship. The derivation graph retains only edges representing genuine derivation, excluding trading protocols, cross-collateral edges, and base-to-base lending edges, while supplementing with CDP minting edges and synthetic edges inferred from token naming conventions. The complete procedure is detailed in Section A.6.

Tier 0 tokens are identified as tokens with zero in-degree in the derivation graph, subject to minimum connectivity and TVL thresholds. This yields 169 base tokens, including native blockchain assets, fiat-backed stablecoins, and major governance tokens. CDP-minted stablecoins are correctly classified as derived tokens, consistent with their position in Mehrling's hierarchy as credit instruments. Tiers propagate via TVL-weighted BFS: $\mathrm{Tier}(B) = \mathrm{Tier}(A) + 1$ for each derivation edge. Sensitivity analysis across 25 parameter configurations confirms that core tokens appear as Tier 0 in all cases, as Section B reports.

## 3.4 Panel Specification and Identification

The panel comprises pool-month observations from February 2022 through December 2025: 2.0 million observations across 4,905 pools after retaining tiers 0–3, requiring non-missing APY and TVL, excluding pools with TVL below $1,000, and winsorizing APY at the 1st and 99th percentiles. Table 2 reports summary statistics by tier.

| Tier | N | Pools | Mean | Median | S.D. | log TVL | Stable % |
|---|---|---|---|---|---|---|---|
| 0 | 560,514 | 1,324 | 7.97 | 1.92 | 14.94 | 12.62 | 33.9 |
| 1 | 1,244,323 | 2,963 | 8.68 | 1.28 | 16.55 | 12.53 | 12.7 |
| 2 | 214,484 | 549 | 6.88 | 2.33 | 12.68 | 13.70 | 17.6 |
| 3 | 17,241 | 69 | 6.64 | 3.30 | 10.94 | 13.28 | 35.3 |
| All | 2,036,562 | 4,905 | 8.28 | 1.58 | 15.72 | 12.68 | 19.2 |

**Table 2** | Descriptive Statistics by Hierarchy Tier. APY in percentage points; Stable % is share of stablecoin-denominated pools.

To test Hypothesis 1, the baseline specification is:

$$\begin{aligned} \text{APY}_{i,t} = \beta \cdot \text{Tier}_i + \gamma \cdot \log\big(\text{TVL}_{i,t}\big) \\ + \delta \cdot \text{Stablecoin}_i + \alpha_t + \varepsilon_{i,t} \end{aligned} \tag{1}$$

where $\alpha_t$ represents time fixed effects and $\text{Stablecoin}_i$ absorbs the base-rate differential between fiat-linked and crypto-native pools. The most saturated specification adds protocol-type and chain fixed effects:

$$\begin{aligned} \text{APY}_{i,t} = \beta \cdot \text{Tier}_i + \gamma \cdot \log\big(\text{TVL}_{i,t}\big) \\ + \delta \cdot \text{Stablecoin}_i + \alpha_t + \mu_p + \nu_c + \varepsilon_{i,t} \end{aligned} \tag{2}$$

Protocol type is plausibly a *mediator* rather than a confounder: the Money View predicts that different intermediary types specialize at different depths, as Table 3 confirms, so controlling for protocol type may absorb the mechanism through which the hierarchy operates.

**Graph distance.** As a complementary measure, the discrete tier variable is replaced with *graph distance*: the shortest-path length in integer hops from each pool's primary token to the nearest Tier 0 token. While tier is a coarse four-level variable, graph distance captures finer-grained structural variation; the Pearson correlation between the two is 0.66, with VIF of 2.4 in single-variable models and 4.0 in the joint specification, confirming moderate collinearity that does not compromise inference. Graph distance is available for 93% of the panel.

**Standard errors** are clustered at the pool level (Petersen 2009). **Identification** rests on saturated controls: protocol-type and chain fixed effects ensure the tier coefficient captures within-category, within-chain variation. Tier assignments are time-invariant, mitigating reverse causality; per-snapshot re-discovery confirms this assumption empirically, as Section C shows. A randomization inference test confirms both channels at conventional significance levels, as Section F reports.

**Event regime analysis.** To test Hypothesis 2, Equation 1 is estimated separately for pre-defined event windows: the Terra collapse, FTX bankruptcy, SVB crisis, and Bitcoin ETF event.

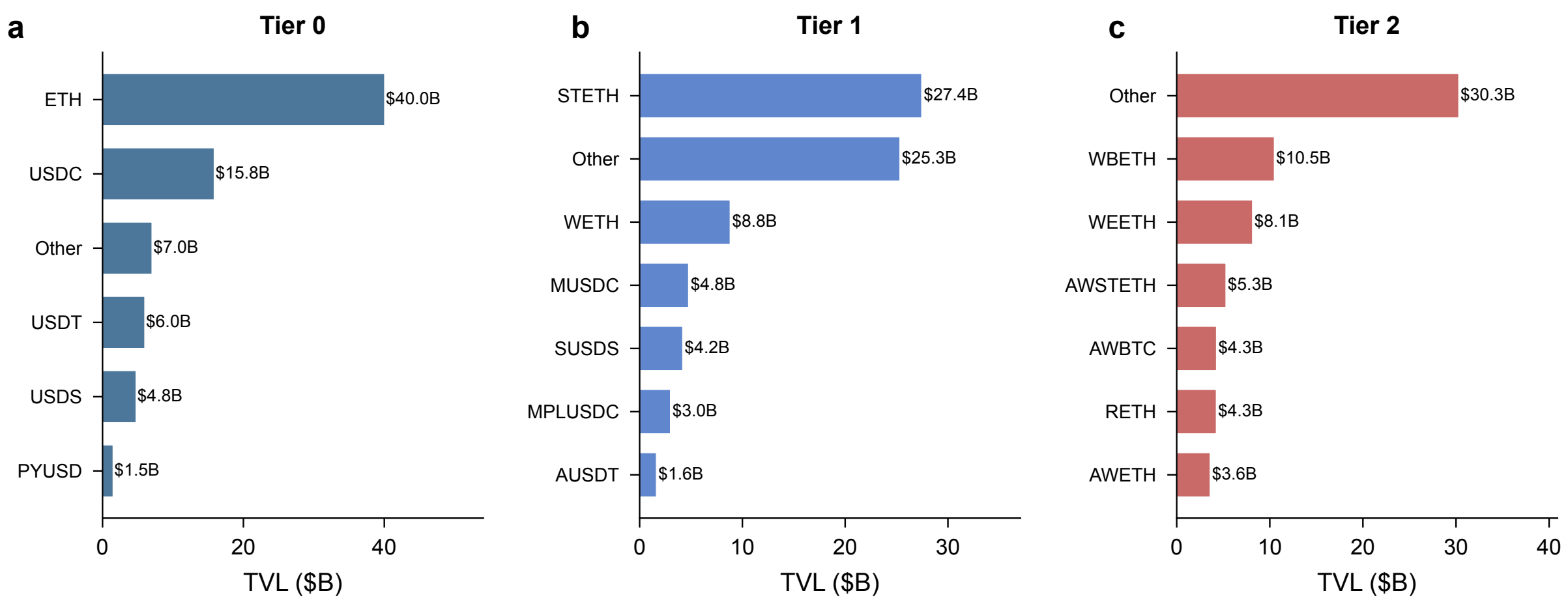

**Fig. 2** | Token Composition by Tier. Panels show the largest tokens by TVL at each tier of the hierarchy: (a) native assets at Tier 0, (b) wrapped and staked derivatives at Tier 1, and (c) second-order derivatives at Tier 2.

# 4 Results

## 4.1 Token Graph Structure

Fig. 2 shows the dominant tokens at each tier. ETH accounts for the majority of Tier 0 value, followed by major stablecoins. At Tier 1, liquid staking derivatives dominate. Tier 2 is headed by Aave receipt tokens and restaking wrappers, reflecting the "stake then lend" pattern examined below. The sharp drop in individual token size combined with a growing residual illustrates both value attrition and increasing fragmentation at higher tiers.

Table 3 decomposes value flows by protocol category, revealing a clear pattern of progressive financialization. At the lowest transition, staking and lending share the role of first-tier intermediary roughly equally. By T2→T3, lending accounts for virtually all flows: staking drops out entirely. This "stake then lend" pattern mirrors the traditional financial system where money-market-like instruments give way to credit intermediation at greater depth. The pattern has a direct empirical consequence: because different protocol types dominate at different tiers, any yield gradient across tiers will partly reflect the changing protocol mix, a compositional effect tested explicitly in Section 4.3.

| Trans. | TVL | Stake | Lend | DEX | Other |
|---|---|---|---|---|---|
| T0→T1 | $72.3B | 40% | 33% | 5% | 22% |
| T1→T2 | $57.1B | 45% | 46% | 3% | 6% |
| T2→T3 | $8.6B | 0% | 99% | 1% | 1% |
| T3→T4 | $5.1B | 0% | 99% | 0% | 1% |

**Table 3** | Protocol Type Distribution by Tier Transition (share of TVL by protocol category).

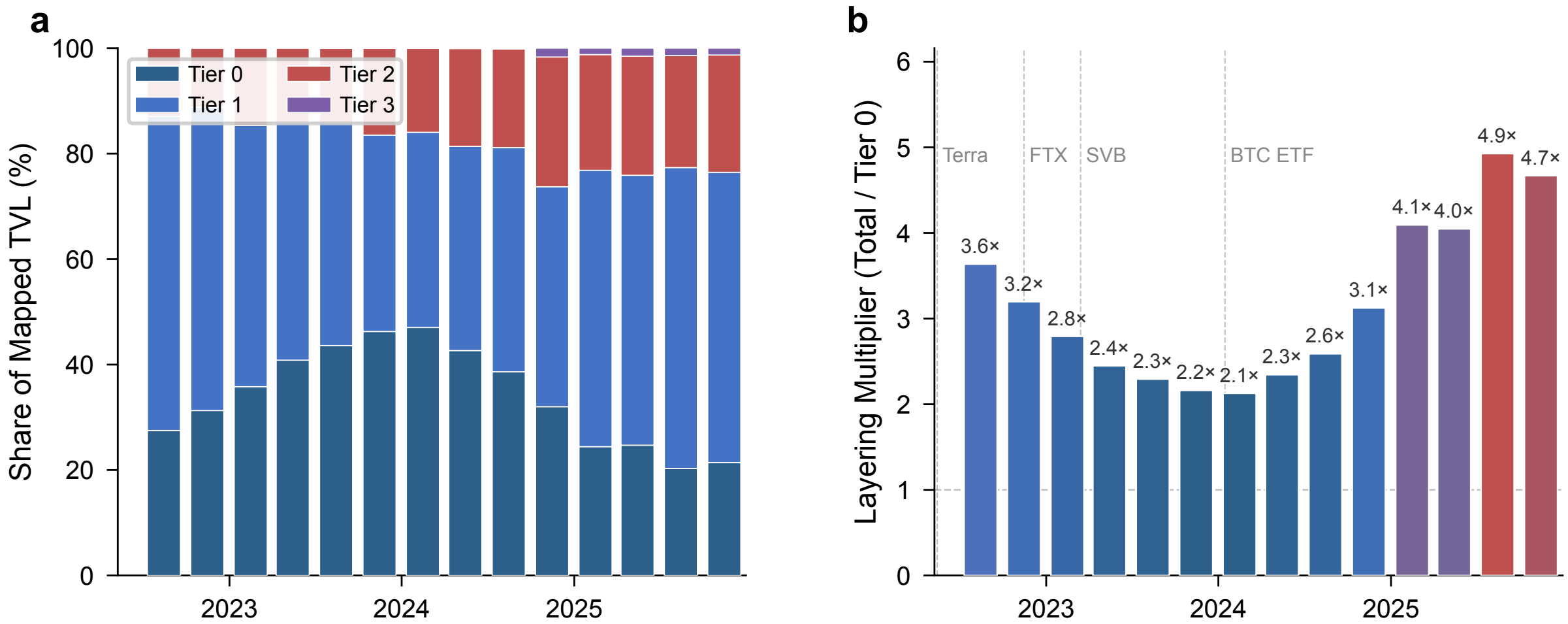

**Fig. 3** | DeFi Hierarchy Composition and Layering Multiplier (2022–2025). (a) Tier shares of mapped TVL. (b) Layering multiplier showing deleveraging through crises followed by re-leveraging to 4.7 × by Q4 2025.

## 4.2 Hierarchy Evolution 2022–2025

Fig. 3 traces the hierarchy through successive crises. Despite absolute TVL volatility, the relative tier distribution remains stable, with Tier 0 and Tier 1 consistently accounting for the vast majority of mapped TVL. Independent per-snapshot discovery yields near-identical assignments from 2023 onward, confirming genuine architectural persistence rather than a static labeling artifact, as Section C details.

The multiplier's excess over unity decomposes by protocol type using edge TVL shares, as detailed in Section A.5. Lending accounts for nearly half the excess, staking for roughly two-fifths, and other protocols for the remainder. Lending and staking together drive over 80% of derivative layering, with lending-driven credit creation, the channel most analogous to traditional deposit multiplication, as the single largest component. The multiplier substantially exceeds the TVL / TVR ratio reported by Luo et al. (2024) the gap reflects scope: Luo et al. measure only the ratio of total to "plain" token TVL, while the present multiplier traces the full derivation chain including staking and restaking layers.

| | (1) Baseline | (2) 1. Prot. FE | (3) 1. Chain FE | (4) 1. Both FE | (5) Excl. LST | (6) Tier Dum. |
|---|---|---|---|---|---|---|
| Tier | $-0.523^{*}$ | $-0.721^{**}$ | 0.008 | $-0.520^{*}$ | $-0.526^{*}$ | |
| | (0.278) | (0.292) | (0.286) | (0.290) | (0.291) | |
| Tier 1 | | | | | | 0.377 |
| | | | | | | (0.484) |
| Tier 2 | | | | | | $-1.768^{***}$ |
| | | | | | | (0.627) |
| Tier 3+ | | | | | | $-0.198$ |
| | | | | | | (1.163) |
| Controls | Yes | Yes | Yes | Yes | Yes | Yes |
| Time FE | Yes | Yes | Yes | Yes | Yes | Yes |
| Protocol-type FE | No | Yes | No | Yes | Yes | Yes |
| Chain FE | No | No | Yes | Yes | Yes | Yes |
| Observations | 2,036,562 | 2,036,562 | 2,036,562 | 2,036,562 | 2,028,900 | 2,036,562 |
| $R^2$ | 0.005 | 0.030 | 0.040 | 0.059 | 0.060 | 0.061 |

**Table 4** | Panel Regression: APY on Hierarchy Tier. Controls: log(TVL), stablecoin indicator. Pool-clustered SEs. $***\, p < 0.01$, $**\, p < 0.05$, $*\, p < 0.1$.

## 4.3 Tier Premium and Embedded Yield Correction

Two complementary measures capture hierarchical position: the discrete tier and graph distance. These tell different stories, and an embedded yield correction is needed to disentangle genuine structural effects from a measurement artifact in DeFi yield reporting.

Table 4 presents the initial puzzle. The baseline gradient is *negative*: pools at deeper tiers report lower APY. Adding protocol-type fixed effects strengthens the negative effect, while chain fixed effects absorb it entirely. The tier dummy specification localizes the pattern to Tier 2 specifically, a non-monotonicity that the embedded yield correction explains. Graph distance deepens the puzzle: each additional hop associates with sharply lower APY, surviving all fixed effects as Section H reports. Yet including *both* tier and graph distance in a joint specification flips the picture: the tier coefficient turns positive while graph distance remains strongly negative. The hierarchy thus encodes two opposing channels, a compositional premium and a structural distance discount, that partially cancel in single-variable models because tier and graph distance are correlated.

**Embedded yield correction.** Platforms such as DefiLlama report only the outermost protocol's APY, omitting accumulated upstream returns. A Tier 2 Aave pool holding weETH reports near-zero lending yield even though the depositor also earns the staking return embedded in weETH. If deeper-tier pools systematically omit more upstream yield, the negative graph distance coefficient could be a reporting artifact.

The correction is constructed as follows. For each hierarchy edge, the creation yield is extracted from DefiLlama's single-token pool data. Each token's total embedded yield is the recursive sum of creation yields from Tier 0 upward: $\text{embedded}(T) = y_{\text{create}(T)} + \text{embedded}(\text{parent}(T))$, with base tokens at zero. Yield data matches for 62% of non-DEX hierarchy edges; unmatched edges default to zero, making the correction conservative. Because the embedded yield is itself estimated from pool data, it introduces a degree of circularity: the correction variable is measured with error from the same data ecosystem as the dependent variable, likely attenuating the corrected coefficients toward zero rather than inflating them.

Table 5 presents the core result. The embedded yield correction attenuates the tier coefficient by 93%, rendering it indistinguishable from zero across all specifications. The graph distance coefficient, by contrast, attenuates by only 23% under full controls, remaining highly significant. In the joint model on corrected APY, both channels emerge cleanly: a compositional tier premium of $+2.0$ pp per tier and a structural distance discount of $-2.9$ pp per hop, both significant at the 1% level.

The resolution is straightforward. The negative tier coefficient is predominantly a measurement artifact: yield underreporting grows with tier depth, so deeper-tier pools appear to earn less when they are in fact reporting less of their total return. The graph distance coefficient survives with only modest attenuation, confirming a genuine structural effect. Controlling for structural distance, each tier carries a compositional premium reflecting lending-dominated upper tiers; controlling for tier, each additional hop carries a structural discount.

| | Original (1) | Corrected (2) | Original (3) | Corrected (4) |
|---|---|---|---|---|
| *Panel A: Tier (full sample)* | | | | |
| Tier | $-0.525^{*}$ | 0.037 | $-0.524^{*}$ | 0.032 |
| | (0.278) | (0.286) | (0.290) | (0.300) |
| *Panel B: Graph Distance* | | | | |
| Graph dist. | $-2.150^{***}$ | $-1.904^{***}$ | $-1.862^{***}$ | $-1.437^{***}$ |
| | (0.303) | (0.312) | (0.390) | (0.401) |
| *Panel C: Joint Model (Full FE)* | | | | |
| Tier | $1.182^{**}$ | $2.016^{***}$ | | |
| | (0.489) | (0.505) | | |
| Graph dist. | $-2.747^{***}$ | $-2.947^{***}$ | | |
| | (0.569) | (0.579) | | |
| Controls | Yes | Yes | Yes | Yes |
| Time FE | Yes | Yes | Yes | Yes |
| Proto + Chain FE | No | No | Yes | Yes |

**Table 5** | Embedded Yield Correction Test. Columns (1, 3): dependent variable is DefiLlama's reported APY. Columns (2, 4): dependent variable is corrected APY (reported + embedded yield). Panel C includes both tier and graph distance. Pool-clustered SEs. $***p < 0.01$, $**p < 0.05$.

## 4.4 Yield Decomposition and Distributional Evidence

Two supplementary analyses clarify where the structural associations operate. First, DefiLlama decomposes pool APY into a *base yield* (the protocol's fundamental return from utilization and fees) and a *reward yield* (incentive token emissions). The tier coefficient concentrates entirely in base yield ($-0.52$ pp, $p < 0.10$; tier dummies: Tier 2 $= -2.46$ pp and Tier 3 $= -2.70$ pp, both $p < 0.01$), while reward yield shows no significant tier gradient. In other words, the hierarchy predicts where fundamental returns decay, but incentive emissions are distributed roughly evenly across tiers. The base-yield regression achieves $R^2 = 6.8\%$, modestly higher than the total-APY specification, suggesting that a substantial share of the "unexplained 94%" reflects reward-token noise orthogonal to hierarchical position.

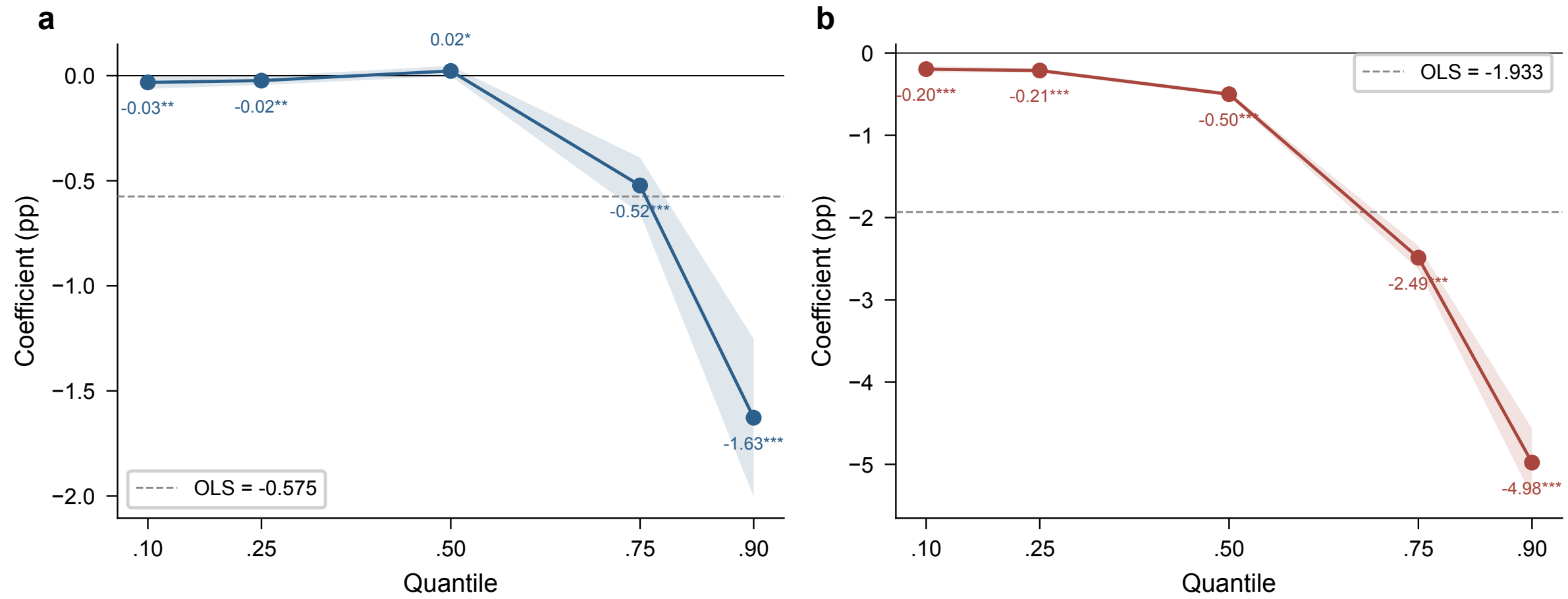

**Fig. 4** | Quantile Regression Coefficients across the yield distribution (Q10 to Q90). (a) Tier coefficient. (b) Graph distance coefficient. The structural associations concentrate in the upper tail: near-zero at the median, strongly negative at Q75–Q90. Dashed lines show OLS estimates; shaded bands show 90% confidence intervals.

Second, quantile regressions reveal that the structural associations are concentrated in the upper tail of the yield distribution. At the median, the tier coefficient is essentially zero; at Q75 it reaches −0.52 pp and at Q90, −1.63 pp, all highly significant. The graph distance coefficient follows the same monotonic pattern, rising from −0.20 pp at Q10 to −4.98 pp at Q90. The OLS estimate of −0.52 pp and 5.9% $R^2$ thus reflects a tail phenomenon: the hierarchy's yield signature is concentrated among high-yield pools, where each derivation step from base assets is associated with sharply lower returns. At the median, where most pools earn modest yields, hierarchical position has little explanatory power.

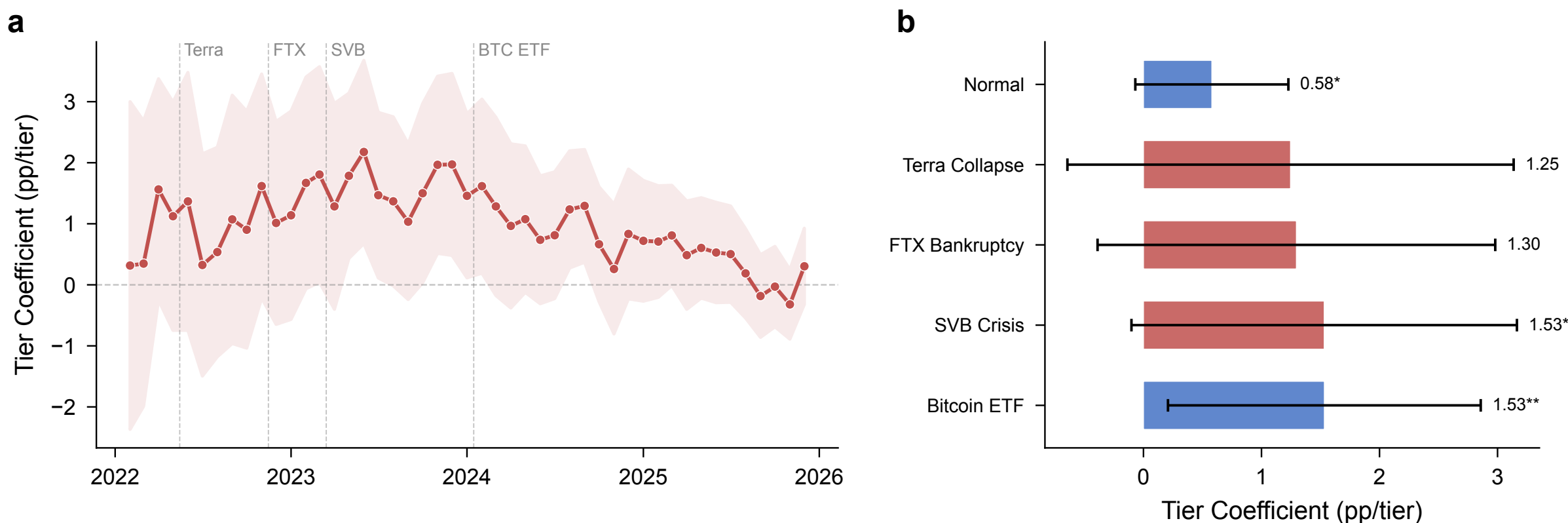

**Fig. 5** | Time-Varying Yield Gradient Across the Token Hierarchy. (a) Monthly tier coefficients from rolling regressions with 95% confidence intervals; vertical lines mark major events. (b) Tier coefficients by market regime.

### 4.5 Crisis Dynamics

Fig. 5 tests whether the yield gradient is time-varying. Of the four event windows examined, only the SVB crisis produces a statistically significant tier-differential effect: the compositional premium roughly triples relative to normal periods. The SVB episode is the one crisis that cascaded mechanically *through* the hierarchy, from USDC's depeg to every derivative built on it (Mehrling 2013). The Terra and FTX episodes show elevated but insignificant estimates; the Bitcoin ETF event is a positive structural shock rather than a crisis and is excluded from the interpretation. With significant evidence from only one of three crises, the support for Hypothesis 2 is suggestive rather than definitive; a longer time series with additional stress episodes is needed to draw firm conclusions about countercyclical repricing.

## 5 Discussion

### 5.1 Liquidity, Not Credit, Drives the Distance Discount

The central finding is that the hierarchy encodes two distinct economic signals previously conflated in raw data. The compositional channel operates through the protocol mix: the "stake then lend" pattern shifts the composition toward higher-yielding lending protocols at deeper tiers, as Table 3 documents. The structural channel operates through derivation distance from base assets and survives the embedded yield correction intact.

Table 6 presents protocol-type subsample regressions that identify the mechanism. Within lending pools alone ($N = 93\{,\}654$, 234 pools), the graph distance coefficient is positive and insignificant ($+1.00$, $p = 0.19$): lending protocols set rates algorithmically based on utilization and risk parameters, regardless of the underlying token's hierarchical position. In DEX pools ($N = 1\{,\}020\{,\}614$), the coefficient amplifies sharply to $-4.41$ pp per hop ($p < 0.001$). The structural distance discount is concentrated in liquidity-sensitive pool types where yield depends on token-pair depth that degrades at greater derivation distance. A direct test using pool-level utilization data confirms this interpretation: mean utilization is roughly flat across tiers (35.6% at Tier 0, 32.7% at Tier 1, 31.4% at Tier 2), and adding utilization to the regression does not attenuate the graph distance coefficient (Sobel $z = -0.27$, $p = 0.79$).

| | (1) Full | (2) Lending | (3) DEX | (4) Liq. Staking | (5) Yield Agg. |
|---|---|---|---|---|---|
| ***Panel I: Graph Distance*** | | | | | |
| Graph dist. | $-1.406^{***}$ | 1.000 | $-4.412^{***}$ | $-5.898^{*}$ | $-0.368$ |
| | (0.325) | (0.760) | (0.726) | (3.263) | (1.566) |
| Controls + FE | Yes | Yes | Yes | Yes | Yes |
| Observations | 1,900,349 | 93,654 | 1,020,614 | 6,396 | 143,107 |
| $R^2$ | 0.034 | 0.139 | 0.053 | 0.306 | 0.108 |
| ***Panel II: Tier*** | | | | | |
| Tier | $-0.098$ | 0.029 | $-0.825^{*}$ | $-1.847^{*}$ | $-0.839$ |
| | (0.286) | (0.189) | (0.501) | (1.073) | (1.131) |
| Controls + FE | Yes | Yes | Yes | Yes | Yes |
| Observations | 2,036,562 | 101,617 | 1,053,148 | 7,662 | 160,733 |
| $R^2$ | 0.037 | 0.129 | 0.039 | 0.278 | 0.108 |

**Table 6** | Protocol-Type Subsample Regressions. Dependent variable: APY (pp). Controls: log(TVL), stablecoin indicator, time FE, chain FE. Pool-clustered SEs. $***\, p < 0.01$, $*\, p < 0.1$.

The hierarchy covaries with risk primarily through the liquidity channel, refining the Money View's prediction from generic "upstream dependency risk" to the specific mechanism of liquidity fragmentation.

This finding carries interpretive weight for the wider literature. Augustin et al. (2022) document "reaching for yield" in DeFi without identifying the structural source; the hierarchy reveals that reaching operates through the compositional channel, as investors move into lending-dominated upper tiers, while the distance discount reflects the cost of doing so in liquidity terms. Par settlement (Mehrling 2011) weakens progressively across tiers: WETH maintains par with ETH through mechanical redemption; stETH through secondary market liquidity; aWETH through overcollateralization and protocol solvency. This progressive weakening aligns with the concentration of the distance discount in pool types where secondary-market liquidity is the binding constraint.

The DeFi hierarchy, despite its aspirations to autonomy, ultimately rests on a foundation extending into traditional banking (Aldasoro et al. 2023). The SVB episode, in which USDC's depeg cascaded mechanically through every derivative built on it, makes this dependence concrete.

## 5.2 Limitations

*Explanatory power.* The full-specification $R^2$ of 5.9% means hierarchical position accounts for a small share of yield variation. The base-yield decomposition ($R^2 = 6.8\%$) and quantile regressions (near-zero at the median, strongly negative at Q90) clarify that the hierarchy is a modest average predictor whose signature concentrates in the upper tail and in fundamental returns, not incentive emissions. The hierarchy is a real structural feature of DeFi, but it is not the dominant determinant of yields; pool-specific factors, reward-token programs, and market-wide conditions collectively drive the bulk of variation.

*Endogeneity.* Despite saturated controls and randomization inference, confounding cannot be fully ruled out. Tier assignments are equilibrium outcomes of protocol design and user behavior; protocols that attract higher-tier tokens may differ systematically in unobservable ways. A fully causal design would require instrumental variables or exogenous shocks to graph structure. The associations documented here should be interpreted as conditional correlations, not causal estimates.

*Data coverage.* Under half of aggregate DeFi TVL is independently verifiable on-chain (Saggese et al. 2025), and off-chain "points" programs are invisible in the data. The embedded yield correction matches 62% of non-DEX edges; characterizing the unmatched 38% reveals they are concentrated in niche protocols with low TVL (1.7% of total edge TVL), not systematically skewed by tier. A bounding exercise imputing unmatched edges at 0% (lower), 50% (middle), and 100% (upper) of tier-average embedded yield shows the graph distance coefficient survives under lower and middle assumptions; the upper bound, which over-corrects by imputing yield to Tier 0 edges with no upstream protocol, attenuates the coefficient toward zero. Reclassifying fiat-backed stablecoins from Tier 0 to Tier 1 (treating them as claims on off-chain reserves) strengthens the tier coefficient ($-0.65$, $p < 0.01$) and preserves the two-channel decomposition (84% tier attenuation), confirming robustness to this alternative classification.

*Protocol-type heterogeneity.* DEX LP positions have concave payoff structures economically distinct from linear lending receipts. The concentration of the distance discount in DEX pools rather than lending pools clarifies its mechanism but also means the full-sample coefficient partly reflects cross-type variation.

*Recursive leverage.* The framework tracks token-type derivations but not within-type quantity leverage from deposit-borrow-redeposit looping, which would require position-level on-chain data.

## 5.3 Policy Implications

The layering multiplier provides a readily computable metric of system-wide derivative layering (Aramonte et al. 2021). Its value for macro-prudential monitoring is qualified, however, by a null predictive result: Granger causality tests ($N = 46$ months) find no evidence that changes in the multiplier predict subsequent TVL drawdowns or yield spread widening ($p > 0.25$ at all lags). The reverse direction is significant ($p = 0.001$ at lag 2), indicating that capital inflows drive multiplier changes rather than the multiplier anticipating stress. This demand-driven dynamic suggests the multiplier reflects current layering depth rather than signaling future fragility, consistent with the traditional banking literature where credit stock is a state variable rather than a leading indicator (Schularick and Taylor 2012). The multiplier is best understood as a contemporaneous structural barometer. For investors, the embedded yield correction has a practical implication: a Tier 3 token's reported yield systematically understates total return, while the structural distance discount means that pools farther from base assets earn genuinely lower marginal yields that widen during crises.

# 6 Conclusion

DeFi's derivative layering is the credit structure, not a measurement artifact. Each dollar of base assets currently supports $4.7 of total claims, with lending and staking driving over 80% of this layering. An embedded yield correction reveals the apparent negative tier gradient as a reporting artifact, attenuated by 93%, and disentangles two channels: a compositional tier premium operating through the protocol mix and a structural distance discount that concentrates in liquidity-sensitive pool types and widens during crises. The Token Graph and layering multiplier provide a step toward the system-level DeFi monitoring that the FSB and IMF have identified as a priority (Financial Stability Board 2023, International Monetary Fund and Financial Stability Board 2023), while the identification of liquidity fragmentation as the primary transmission channel refines the Money View's application to decentralized markets.

## Acknowledgements

The author thanks Rashad Ahmed, Nicola Gennaioli, Campbell Harvey, Simon Mayer, Dirk Niepelt, and Andreas Park for helpful comments and suggestions, and Patrick Scott and the DefiLlama team for data access. All errors are the author's own.

## Declarations

- **Funding:** This research received no specific grant from any funding agency in the public, commercial, or not-for-profit sectors.
- **Conflict of interest:** The author declares no competing interests.
- **Data availability:** The datasets analyzed during the current study are available from DefiLlama (https://defillama.com). Processed data and replication code are available from the corresponding author upon reasonable request.
- **Code availability:** Replication code is available from the corresponding author upon reasonable request.

- **CRediT author statement:** Wenbin Wu: Conceptualization, Methodology, Software, Formal analysis, Data curation, Visualization, Writing – original draft, Writing – review & editing.
- **Declaration of generative AI in the writing process:** During the preparation of this work the author used generative AI tools in order to improve language and readability. After using these tools, the author reviewed and edited the content as needed and takes full responsibility for the content of the publication.

# A Formal Definitions

## A.1 Token Contracts and Operations

Let $\mathcal{X}$ denote the set of all tokens in the DeFi ecosystem, and let $\mathcal{P}$ denote the set of all protocols. For each protocol $P_i \in \mathcal{P}$:

- $X_L(P_i) \subseteq \mathcal{X}$: the set of tokens that protocol $P_i$ accepts as inputs
- $X_G(P_i) \subseteq \mathcal{X}$: the set of tokens that protocol $P_i$ generates as outputs
- $X_C(P_i, t) \subseteq X_L(P_i)$: the set of tokens currently locked in protocol $P_i$ at time $t$

The Total Value Locked for protocol $P_i$ at time $t$ is:

$$\text{TVL}(P_i, t) = \sum_{x \in X_C(P_i, t)} q_i(x, t) \cdot v(x, t) \tag{3}$$

where $q_i(x, t)$ is the quantity of token $x$ held by protocol $P_i$ at time $t$, and $v(x, t)$ is the USD price of token $x$ at time $t$.

## A.2 Token Graph and Derivation Graph

The *full token graph* represents all token transformation relationships across the ecosystem:

$$G_{\text{full}} = (\mathcal{X}, \mathcal{E}_{\text{full}}) \tag{4}$$

where $\mathcal{E}_{\text{full}} \subseteq \mathcal{X} \times \mathcal{X} \times \mathcal{P}$ is the set of directed edges. An edge $(x_1, x_2, P_i) \in \mathcal{E}_{\text{full}}$ indicates that protocol $P_i$ accepts token $x_1$ and generates token $x_2$.

The *derivation graph* $G_{\text{deriv}} = (\mathcal{X}, \mathcal{E}_{\text{deriv}})$ is obtained from $G_{\text{full}}$ by removing trading protocols, cross-collateral edges, and base-to-base lending edges, and adding CDP minting edges and synthetic name-containment edges, as detailed in Section 3 and Section A.6. Tier assignment operates on $G_{\text{deriv}}$.

## A.3 Tier Assignment

Let $\mathcal{X}_{\text{base}} \subset \mathcal{X}$ denote the set of base assets. The tier of a token is the length of the shortest directed path from any base asset in $G_{\text{deriv}}$:

$$\text{Tier}(x) = \begin{cases} 0 & \text{if } x \in \mathcal{X}_{\text{base}} \\ 1 + \min_{e \in \text{In}(x)} \text{Tier}(\text{src}(e)) & \text{if } \text{In}(x) \neq \varnothing \\ -1 & \text{otherwise (unmapped)} \end{cases} \tag{5}$$

where $\mathrm{src}(e)$ denotes the source token of edge $e$. The minimization selects the parent with the lowest tier; this is equivalent to BFS from $\mathcal{X}_{\mathrm{base}}$ through $G_{\mathrm{deriv}}$, as implemented in Section A.6.

## A.4 Layering Multiplier

The system-wide layering multiplier at time $t$ is:

$$\mathrm{LM}(t) = \mathrm{TVL}_{\mathrm{mapped}} \frac{t}{\mathrm{TVL}_{\mathrm{Tier}\ 0}}(t) \tag{6}$$

where $\mathrm{TVL}_{\mathrm{mapped}(t)}$ is total mapped TVL across all protocols and $\mathrm{TVL}_{\mathrm{Tier}\ 0}(t)$ is the TVL attributable to Tier 0 tokens. The multiplier captures all forms of derivative layering and is thus broader than the traditional banking money multiplier.

## A.5 Multiplier Decomposition by Protocol Type

The excess layering multiplier is allocated to protocol types proportionally to their share of tier-increasing edge TVL. For each tier-increasing edge $(x, y) \in \mathcal{E}_{\mathrm{deriv}}$ with $\mathrm{Tier}(y) = \mathrm{Tier}(x) + 1$, the mediating protocol is classified into category $k \in \{\mathrm{Lending}, \mathrm{Staking}, \mathrm{DEX}, \mathrm{Other}\}$:

$$w_k(t) = \frac{\sum_{(x,y):\ \mathrm{type}(P_{x \to y}) = k} \mathrm{TVL}_{\mathrm{edge}}(x, y, t)}{\sum_{(x,y)} \mathrm{TVL}_{\mathrm{edge}}(x, y, t)} \tag{7}$$

The protocol-type contribution is then $\Delta\mathrm{LM}_k(t) = w_k(t) \cdot (\mathrm{LM}(t) - 1)$, so that $\sum_k \Delta\mathrm{LM}_k = \mathrm{LM} - 1$. For multi-hop paths, each hop contributes independently to its respective protocol type's edge TVL share.

## A.6 Tier Assignment Algorithm

The complete procedure for discovering the token hierarchy from pool-level data:

**Algorithm 1:** Token Hierarchy Discovery

---

**Input:** Pool data $\mathcal{D}$ with input/output token mappings; protocol categories $\mathcal{C}$

**Output:** Tier assignment $\tau : \mathcal{X} \to \{0, 1, 2, ...\} \cup \{-1\}$

---

**1. Build Full Graph.** For each pool $p \in \mathcal{D}$, for each valid token pair $(x_{\text{in}}, x_{\text{out}})$ with $x_{\text{in}} \neq x_{\text{out}}$, add edge $(x_{\text{in}}, x_{\text{out}})$ with TVL weight to graph $G_{\text{full}}$.

**2. Build Derivation Graph.** Starting from $G_{\text{full}}$:

(a) remove edges from trading protocols via $\mathcal{C}$

(b) filter cross-collateral edges: if output token appears among inputs, suppress edges to it

(c) remove base-to-base lending edges where both tokens are significant base assets

(d) for wrapping pairs, reverse to creation direction using known DeFi prefixes

(e) apply Tier 0 protection: remove non-name-related incoming edges to strong base-token candidates

(f) add explicit CDP minting edges for protocols absent from yield data

(g) add synthetic edges via composite name parsing, prefix stripping, and longest-substring matching

Call the result $G_{\text{deriv}}$.

**3. Discover Tier 0 Tokens.** Set $\mathcal{X}_0 = \left\{x \in G_{\text{deriv}} : \text{in-deg}_{\text{deriv}(x)} = 0 \text{ and out-deg}_{\text{full}(x)} \geq 3 \text{ and TVL}_{\text{src}(x)} \geq 1\text{M}\right\}$. For pairs $(a, b) \in \mathcal{X}_0$ where $a$ is a prefix or suffix of $b$ and $\text{TVL}(a) \geq \text{TVL}(b)/1000$, demote $b$ to Tier 1 under parent $a$.

**4. Propagate Tiers via TVL-weighted BFS.** Set $\tau(x) = 0$ for all $x \in \mathcal{X}_0$. Initialize priority queue $Q$ with successors of $\mathcal{X}_0$ in $G_{\text{deriv}}$, ordered by descending TVL.

**while** $Q \neq \varnothing$ **do**

Dequeue highest-TVL token $y$ with parent $x$

**if** $\tau(y)$ undefined **then** set $\tau(y) = \tau(x) + 1$ and enqueue successors of $y$

Set $\tau(x) = -1$ for all unvisited $x$.

**5. Return** $\tau$.

| Min. out-deg | $0 | $100K | $1M | $10M | $100M |
|---|---|---|---|---|---|
| 2 | 0.75 | 0.80 | 0.85 | 0.55 | 0.18 |
| 3 (base) | 0.80 | 0.88 | **1.00** | 0.61 | 0.20 |
| 5 | 0.68 | 0.73 | 0.82 | 0.52 | 0.17 |
| 7 | 0.55 | 0.60 | 0.68 | 0.43 | 0.15 |
| 10 | 0.40 | 0.44 | 0.51 | 0.33 | 0.12 |

**Table 7** | Jaccard Similarity of Tier 0 Token Set vs. Baseline (out-degree $\geq$ 3, TVL $\geq$ $1M).

### A.7 Circular Dependencies

The 79 cycles identified in Section 3 arise predominantly from bidirectional lending. Excluding the trivial ETH $\leftrightarrow$ WETH wrapper, cycle-involved TVL is negligible. BFS resolves these by assigning the minimum tier from the shortest path to Tier 0.

## B Tier Assignment Sensitivity

The minimum out-degree and TVL threshold are varied across 25 configurations and Jaccard similarity computed against the baseline.

Core tokens appear as Tier 0 in all 25 configurations. For nearby parameter choices, Jaccard similarity exceeds 0.80. Sensitivity increases at extremes, though even extreme configurations retain core tokens. Reclassifying economically ambiguous tokens between Tier 0 and Tier 1 leaves the main tier coefficient qualitatively stable.

## C Temporal Stability of Tier Assignments

The baseline uses time-invariant tier assignments from the full-sample token graph. To verify that this does not impose artificial stability, the tier discovery algorithm is rerun independently on each quarterly snapshot.

From Q1 2023 onward, covering 88% of the panel, per-snapshot tier assignments agree with the full-sample baseline for over 95% of tokens. All core tokens receive identical tier assignments in every quarter they appear. Early quarters show lower agreement due to sparse data coverage rather than genuine structural instability, confirming that the time-invariant assumption is empirically justified.

| Quarter | Pools | Tier 0 | Agreement | Spearman $\rho$ |
|---|---|---|---|---|
| 2022Q1 | 1,085 | 11 | 63.3% | 0.715 |
| 2022Q3 | 1,905 | 29 | 91.6% | 0.911 |
| 2023Q1 | 2,413 | 34 | 95.4% | 0.948 |
| 2024Q1 | 4,569 | 66 | 96.9% | 0.967 |
| 2025Q1 | 8,293 | 104 | 96.2% | 0.955 |
| 2025Q4 | 14,252 | 168 | 100.0% | 1.000 |

**Table 8** | Temporal Stability of Tier Assignments. Agreement: percentage of common tokens receiving the same tier as the full-sample baseline. Spearman $\rho$: rank correlation of tier assignments for common tokens.

| Outlier Treatment | Tier Coeff. | SE | $R^2$ |
|---|---|---|---|
| Baseline (APY < 100) | −0.520 | (0.290) | 0.059 |
| Winsorize 1st/99th | −0.504 | (0.285) | 0.061 |
| Winsorize 2nd/98th | −0.475 | (0.276) | 0.064 |
| Winsorize 5th/95th | −0.362 | (0.237) | 0.074 |
| Trimmed 1st/99th | −0.407 | (0.255) | 0.061 |

**Table 9** | Tier Coefficient Under Alternative Outlier Treatments (Column 4 specification). $N = 2\{,\}036\{,\}562$ for winsorization; $N = 2\{,\}016\{,\}197$ for trimming.

# D Winsorization Robustness

The baseline specification filters APY to $[0, 100)$. Table 9 reports the tier coefficient under alternative outlier treatments.

The coefficient ranges from $-0.36$ to $-0.52$ pp across all treatments, confirming the negative tier-yield association is not driven by extreme observations.

# E Survivorship Bias

Only 75 of 4,905 pools (1.5%) exit before the observation period ends. Exit rates are uniformly low across tiers, with no significant correlation between tier and exit ($p = 0.12$). Table 10 restricts to pools surviving at least 6 or 12 months.

Long-lived and medium-lived subsamples are consistent with the full-sample estimate. Survivorship bias does not materially affect results.

| Sample | Pools | Tier Coeff. | SE | $R^2$ |
|---|---|---|---|---|
| Full sample | 4,905 | −0.520 | (0.290) | 0.059 |
| Long-lived (≥ 12 mo) | 2,487 | −0.480 | (0.331) | 0.056 |
| Medium-lived (≥ 6 mo) | 3,441 | −0.546 | (0.300) | 0.059 |
| Spanning FTX crisis | 651 | −0.325 | (0.603) | 0.034 |

**Table 10** | Survivorship Robustness: Tier Coefficient Across Balanced Panel Definitions. All use the full-specification (Column 4).

# F Placebo Test

Randomization inference permuting tier labels across pools 1,000 times (Hagemann 2019, Eggers et al. 2024) places the actual baseline coefficient at the 4.4th percentile of the permutation distribution (one-sided $p = 0.044$). For graph distance under full fixed effects, the actual coefficient is highly significant ($p < 0.001$). Both channels survive at conventional significance levels.

# G Oster (2019) Coefficient Stability

Under the proportional-selection assumption with $R_{\text{max}} = 1.3R_{\text{full}}$, both tier and graph distance coefficients exhibit $\delta^* < 0$: adding controls strengthens rather than attenuates the estimates (Oster 2019). As Diegert et al. (2022) note, negative $\delta^*$ provides weaker identification guarantees; the result indicates that the observable control set does not inflate the estimates rather than serving as a definitive robustness certificate.

# H Graph Distance Regressions

Each hop associates with substantially lower APY across all specifications. Chain fixed effects produce the largest attenuation, suggesting some of the graph distance effect reflects chain-level differences. The joint specification confirms the two opposing channels identified in Section 4.3.

# I Algorithm Robustness: Leave-One-Step-Out

Removing each filtering step individually and comparing the resulting hierarchy to the baseline reveals that each step is either structurally necessary or empirically inconsequential. The nine core tokens retain their baseline tier in four of six step removals. Removing the trading-protocol filter floods the graph with DEX tokens but does not affect core assignments; removing edge reversal severs the ETH derivation chain entirely. The remaining steps affect fewer than 8% of tokens each.

| | (1) Baseline | (2) 1. Prot. FE | (3) 1. Chain FE | (4) 1. Both FE | (5) Joint |
|---|---|---|---|---|---|
| Graph dist. | $-2.150^{***}$ | $-2.169^{***}$ | $-1.305^{***}$ | $-1.862^{***}$ | $-2.747^{***}$ |
| | (0.303) | (0.388) | (0.325) | (0.390) | (0.569) |
| Tier | | | | | $1.182^{**}$ |
| | | | | | (0.489) |
| Controls | Yes | Yes | Yes | Yes | Yes |
| Time FE | Yes | Yes | Yes | Yes | Yes |
| Protocol-type FE | No | Yes | No | Yes | Yes |
| Chain FE | No | No | Yes | Yes | Yes |
| Observations | 1,900,190 | 1,900,190 | 1,900,190 | 1,900,190 | 1,900,190 |
| $R^2$ | 0.012 | 0.034 | 0.036 | 0.059 | 0.059 |

**Table 11** | Panel Regression: APY on Graph Distance. Controls: log(TVL), stablecoin indicator. Sample restricted to pools with mapped graph distance (93% of full panel). Column (5) includes both tier and graph distance. Pool-clustered SEs. $***\,p < 0.01$, $**\,p < 0.05$.

# J Crisis Robustness

A daily panel confirms all three crises produce significant APY spikes in tight ±30-day event windows. A placebo test with 1,000 random pseudo-crisis dates validates Terra and SVB as genuine outlier events. The tier-by-crisis interaction is not significant at daily frequency, suggesting the immediate shock propagates uniformly across tiers; the tier-differential repricing documented in Fig. 5 emerges over subsequent weeks as markets adjust.